# Ferroelectric imprint in annealed $Bi_{0.9}La_{0.1}Fe_{0.9}Mn_{0.1}O_3$ thin films

T. T. Carvalho [1,4], F. G. Figueiras [1,2], M. P. F. Garcia [1], A. M. Pereira [1], J. R. A. Fernandes [3], J. Perez de la Cruz [1], S. M. S. Pereira[2], P. B. Tavares [4], A. Almeida [1], J. Agostinho Moreira [1]

1- IFIMUP-IN, Physics and Astronomy Department, Faculty of Sciences, University of Porto, R. Campo Alegre, 687, 4169-007 Porto, Portugal.
2- CICECO-AIM and Physics Departament, University of Aveiro, 3810-193 Aveiro, Portugal.
3- INESC TEC and Physics Departament, University of Trás-os-Montes e Alto Douro, 5001-801 Vila Real, Portugal.
4- CQVR and Chemistry Department, University of Trás-os-Montes and Alto Douro, 5001-801 Vila Real, Portugal.

**Abstract**

The present work reports the study of the optimized processing conditions of $Bi_{0.9}La_{0.1}Fe_{0.9}Mn_{0.1}O_3$ thin films, grown by RF sputtering on platinum metalized silicon substrates. The combination of deposition at relatively low substrate temperature followed by adequate ex situ annealing leads to thin films with smooth surface morphology and the formation of a high-quality monophasic layer, with the $(100)_c$ preferable orientation. The annealed films show ferroelectric imprint.

**Introduction**

The potential use of the multiferroic $BiFeO_3$ (BFO) thin film represents an advanced step for the development of a new generation of micro and nanoelectronic devices [1]. However, the formation of secondary phases, which increases the leakage current, limits some applications of the BFO films [2]. In this context, it has been reported that partial substitution of Bi ion by rare earth ions (such as La, Pr, Nd) or/and of iron by transition metal ions (like Ti, Cr, Mn) reduces the leakage current and enhances the ferroelectric properties [3,4]. Previous studies have shown that the substitution of bismuth by lanthanum inhibits the formation of secondary phases [5], while the introduction of manganese in the Fe-site has been used in order to suppress the valence fluctuation of the $Fe^{3+}$ ions [4]. For instance, studies on $Bi_{0.9}La_{0.1}Fe_{0.9}Mn_{0.1}O_3$ (BLFM) revealed absence of secondary phases, ferromagnetic behavior and low leakage current, contrarily to undoped $BiFeO_3$ [6, 7]. However, beyond the controversy involving the existence of ferroelectricity, single-phase BFO thin films have a narrow processing window, making the control of substrate temperature and oxygen partial pressure crucial deposition parameters [8, 9]. It is known that at relative low substrate temperature (~550 ºC) or high oxygen pressure ($10^{-1}$ mbar) $Bi_2O_3$ is formed, while at high temperatures (>650ºC) or low oxygen pressure ($10^{-3}$ mbar), a $Fe_2O_3$ phase has been found [8, 9]. In BFO thin films, the high leakage current has been attributed to oxygen and bismuth vacancies [10, 11].

Different attempts have been undertaken to reduce the oxygen vacancies, namely by using different deposition oxygen pressures [12] and annealing treatment in oxygen rich atmosphere [10, 13].

However, the processing route to obtain high quality monophasic BFO and doped BFO thin films has not been optimized yet.

Furthermore, the effect of the post annealing process has been studied in BFO and doped BFO thin films, revealing that this process is most interesting to control the domain pattern in ferroelectric films [14].

This work aims at studying the optimal deposition and post-annealing conditions towards the formation of high quality monophasic BLFM films with enhanced ferroelectric response.

**Experimental Methods**

Thin films were deposited by RF sputtering on metalized platinum substrates (Pt/Ti/SiO$_2$/Si(100)), using an *US GUN II 2''* sputter source with an applied RF power of 100 W. For this purpose, a Bi$_{0.9}$La$_{0.1}$Fe$_{0.9}$Mn$_{0.1}$O$_3$ ceramic target, with 5 cm in diameter, was synthesized by sol gel combustion method, and fully characterized, as previously reported [15]. The films were deposited during 30 minutes on 400 ºC to 650 ºC heated substrates, using a 9:1 Argon/Oxygen ratio at 1.2x10$^{-2}$ mbar pressure in the deposition chamber. Films deposited at 400 ºC were further post-annealed in air at 575ºC during a period of 30 minutes with fast cooling rate. Systematic characterization of the as processed films was performed in order to determine the effect of the deposition conditions on the structure, morphology and chemical composition of the films. X-ray diffraction data were collected at room temperature in a *PANalytical X'Pert Pro* diffractometer, equipped with *X'Celerator* detector and secondary monochromator. The measurements were carried out in θ/2θ Bragg-Bentano geometry, using CuKα radiation (λ=0.15418 nm), from 20º to 60º, at 0.017º step width and 150 s/step of counting time. For the crystalline orientation study or in the pole-figure measurement, the diffracted beam intensity is measured in a *c* range from 0° to 80°, while keeping fixed the position of the detector at a specific reflection angle 2θ. Pole densities are plotted in stereographic projection with polar angle c. The plane of the stereographic projection is chosen parallel to the sample surface. Unpolarized micro-Raman spectra were recorded at room temperature by using a *T64000* spectrometer, and 514.5 nm as excitation, in the spectral range 200 - 850 cm$^{-1}$. The laser power at the sample was 5 mW, in order to avoid sample self-heating. The morphology of the films were analyzed by SEM with a *FEI Quanta 400/EDAX*. Local piezoelectric response of the films surface was investigated by PFM using a *NT-MDT Ntegra* equipped with a lock-in amplifier (*SR-830A Stanford Research*), a function generator (*FG-120, Yokagawa*) and appropriate *Nanosensors PPP-NCHR* cantilevers. Bias lithography encompass regions were stimulated at +- 20 V$_{d.c.}$ and 1 Hz line scan rate. Low dc field induced specific magnetization and magnetic hysteresis loops measurements were carried out using a commercial superconducting quantum interference device SQUID.

## Results and discussion

### A. Substrate temperature

The first step of this work was the study the effect of the substrate temperature on the stoichiometry and crystal structure of the as-prepared BLFM thin films. Their morphology was analysed by SEM. A representative cross section SEM image of film deposited at 550ºC is shown in Figure 1, and reveals a smooth film surface and an average thickness of 120 nm. Similar results were obtained for the other films.

Figure 2a shows the XRD patterns of the $Bi_{0.9}La_{0.1}Fe_{0.9}Mn_{0.1}O_3$ target and of the films deposited at fixed substrate temperatures, between 400 °C to 600 °C. The XRD patterns of the films deposited at substrate temperatures below 550 °C only exhibit the Bragg peaks arising from the substrate, evidencing that these films are amorphous, while for the films deposited at higher substrate temperature, well defined Bragg peaks are observed, pointing out to the formation of crystalline phases. However, as the substrate temperature increases from 625 °C to 650 °C, the intensity of the Bragg peaks decreases. We will address to this issue later on. The Bragg peaks were indexed according to a pseudocubic unit cell [10], and the pseudocubic lattice parameters of the films processed at substrate temperatures T ≥ 575 °C were calculated, as well as of the bulk target. The results are shown in Table 1. Relatively to the bulk values, the $a_{pc}$ increases about 3-4%, while both $b_{pc}$ and $c_{pc}$ decrease 1% as the substrate temperature increases. Due to the rather large change of $a_{pc}$, compared with the parameters in the other two directions, the unit cell volume increases with the substrate temperature.

Figure 2b shows the unpolarized Raman spectra of the bulk target and the films deposited at several fixed substrate temperatures. The Raman spectrum of the bulk target exhibits several bands below 400 cm$^{-1}$, which are assigned to lattice vibrations involving the $Bi^{3+}$ and $La^{3+}$ ions [16]. As we can see from Figure 2b, these bands are rather intense in the Raman spectra of films deposited at substrate temperatures T ≤ 600 °C, and very weak in the Raman spectrum of the film deposited at 650 °C. The decrease of intensity of vibrational bands associated with A-site cation motion points to a volatilization of $Bi^{3+}$, which, in turn, suggests the presence of spurious iron oxide phases in this film, most likely hematite. Furthermore, sharp Raman bands, localized at 222, 289 and 406 cm$^{-1}$ (signalized by * in Figure 2b), are observed. These bands correspond to $A_{1g}$ and E modes of hematite, $Fe_2O_3$ [17, 18]. The existence of hematite was confirmed by M(H) measurements (see Figure 2c), which clearly shows an increase of the magnetic response of the film deposited at 650 °C relatively to any other film deposited at lower temperatures due to the canted antiferromagnetic or weakly ferromagnetic state of hematite. The volatilization of $Bi^{3+}$ was also evidenced by the EDS analysis of the films. Bi/Fe ratio is about 1 for the films deposited at the lowest temperatures, and decreases dramatically as the substrate temperature increases, taking the value of 0.5 for the film deposited at 650 °C. The volatilization of $Bi^{3+}$ creates defects and disorder in the crystal lattice which explains the decrease of intensity of the Bragg peaks of the XRD pattern and the Raman bands at low frequencies ($\omega < 400$ cm$^{-1}$) of the Raman spectrum recorded in this film.

The strong and broad band located at around 630 cm$^{-1}$ is assigned to the symmetric stretching mode of the MnO6 octahedra [19]. The existence of this band, observed in the Raman spectra recorded in the bulk sample, and the films deposited at temperatures T≥ 550 °C, evidences the incorporation of $Mn^{3+}$ ion in the B-site of the lattice. The symmetric stretching mode is known to be very sensitive to disorder and strain [20]. We have determined the wavenumber and line width of the corresponding Raman band. The results are presented in Table 2. The line width of the MnO6 symmetric stretching mode is about 71±2 cm$^{-1}$ in the bulk and, then, increases 14% in the thin films, being almost constant in the deposition temperature range between 550 – 600 °C. This value becomes 40% larger, relatively to the bulk value, when the deposition temperature is 650 °C. The increase of the line width of the symmetric stretching mode points for structural disorder in the films, which increases as the deposition temperature increases. The frequency of the symmetric stretching mode is practically unchanged from bulk to films deposited for T≤600 °C, but increases about 7 cm$^{-1}$ in the film deposited at 650 °C, meaning that the Mn-O bond length of the MnO6 octahedra decreases about 0.6% relatively to the bulk value, although the unit cell volume expands. This result points, once again, for distortions arising from disorder and defects due to $Bi^{3+}$ volatilization.

Summarizing the main results presented in this section, the experimental results clearly show that the films processed at low temperatures, 550 °C under an atmosphere of 1.2×10$^{-2}$ mbar, exhibit the stoichiometric BLFM phase, but rather poor crystalline structure.

**B. Post annealing effect**

In the second step of this work, we have studied the effect of post-annealing on the crystalline and polar properties of BLFM thin films deposited at low temperatures, where the crystalline phase is absent. For this study, the film deposited at 400 °C was chosen, which underwent after deposition post-annealing in air at 575 °C during 30 minutes. Figure 3a shows the XRD pattern of the film after post annealing, along with the XRD pattern of the as-processed film at 600 °C. Comparing the XRD patterns of the film deposited at 400 °C before (see Figure 2a) and after annealing, it becomes evident that the annealing process improved the crystallinity of the film. The enhancement of the film crystallinity and the absence of spurious phases are further corroborated through the analysis of the Raman spectrum, shown in Figure 3b. The Raman spectrum of the post-annealed film is similar to the bulk sample, excluding the formation of spurious phases in the post-annealing process. The lattice modes below 400 cm$^{-1}$ are well resolved, and the line width of the symmetric stretching mode is 68±2 cm$^{-1}$, which is similar to the bulk sample (see Table 2). The ensemble of these results fully confirms the enhanced crystallization of the film due to ex-situ annealing at 575 °C. Moreover, comparing the XRD patterns presented in Figure 3a, we observe an intensity decrease of the Bragg peaks indexed to the (110) and (211) planes, evidencing the (100)c preferential orientation of the post-annealed film. The preferable orientation could be associated with the formation of a $Bi_2O_3$ layer at the interface of Pt/BLFM as it has been observed in $BiFeO_3$ thin films [7, 21]. It has been referred to that this layer acts as a Bi source for compensating Bi volatilization, and as a diffusion barrier for species $BiFeO_3$ [7, 21]. A similar process can take place in our case.

The epitaxy and preferential orientation growth of the post-annealed film can be confirmed from the Poles figure, shown in Figure 3c. As it can be seen, the poles figures exhibit four prolonged lobes aligned with the substrate edges, which are indexed to $(100)_C$ reflections, evidencing both epitaxial and relaxation effects in the film layer.

Once demonstrated the improvement of the crystalline and chemical quality of the post-annealed film, we analyzed the local ferroelectric properties of the films. Scanning force microscopy studies based in piezo-response mode (PFM) allowed observing both the topographic features and the piezo-response properties of the films. Figure 4 shows the PFM images recorded in the post-annealed film. The surface analysis shows an generally homogeneous morphology, with profile displaying grain size spreading close to 100 nm and off-plane roughness below 20 nm.

The as-deposited film at 575°C shows no domain contrast and no relevant imprint even after performing ±20 $V_{d.c.}$ bias poling lithography (not show). This passive response enables to settle it as a control sample.

The post-annealed film exhibits a domain structure which can be clearly observed in figure 4, without any correlation to topography. The existence of coherent domains profiles spreading up to ~1 μm size without topographical cross-talk is evident. Figure 4 shows PFM contrast on amplitude and phase after poling the films with ±10 V voltage, confirming strong ferroelectric imprint and clear domain reversal associated with bias lithography sequence, without any relevant topographic changes.

**Conclusions**

This work reports on the effect of the ex-situ post-annealing of $Bi_{0.9}La_{0.1}Fe_{0.9}Mn_{0.1}O_3$ thin films deposited at low temperatures. It was clearly demonstrated that the deposition at low temperatures (T < 550°C), although giving rise to amorphous films, the stoichiometry of the target is preserved. A post-annealing of the amorphous films, in air at T = 575 °C during 30 minutes, induces both crystallinity and (100)c preferential growth of the phase, preserving its stoichiometry, and preventing defects and disorder to emerge, It was also revealed that the ferroelectric response of the post-annealed film is improved relatively to the as-processed film at 575 °C, enabling electric domain writing on the film. This property is very attractive for applications, as in the case of four state memories.

**Acknowledgements**

This work was developed in the scope of the project Norte-070124-FEDER-000070, I3N (PEst-C/CTM/LA0025/2013-14), CICECO-Aveiro Institute of Materials (UID/CTM/50011/2013), and PTDC/FIS-NAN/0533/2012, grant SFRH/BD/41331/2007; SFRH/BPD/80663/2011 financed by national funds through the FCT/MEC, and when applicable co-financed by FEDER under the PT2020 Partnership Agreement.

Table 1 Lattice parameters and volume of the cell for the bulk and as-grown films at different temperatures.

| sample | a/√2 (Å) | b/2 (Å) | c/√2 (Å) | V (Å³) |
|---|---|---|---|---|
| Bulk | 3.86(4) | 3.95(9) | 3.95(6) | 60.5 |
| 575ºC | 3.97(8) | 3.91(3) | 3.92(7) | 61.1 |
| 600ºC | 4.02(2) | 3.92(3) | 3.93(0) | 62.0 |
| 625ºC | 4.00(5) | 3.92(3) | 3.92(9) | 61.7 |
| 650ºC | 4.03(2) | 3.91(1) | 3.91(9) | 61.8 |

Table 2 Wavenumber and line width of the symmetric stretching mode of the MnO6 octahedra for bulk, as-grown films at different temperatures, and as-prepared film at 400ºC, post annealed at 575ºC

| sample | wavenumber (cm$^{-1}$) | Line width (cm$^{-1}$) |
|---|---|---|
| Bulk | 627±2 | 71±2 |
| As grown 550ºC | 630±2 | 84±3 |
| As grown 600ºC | 627±1 | 83±3 |
| As grown 650ºC | 633±2 | 101±4 |
| Post annealed 575ºC | 634±1 | 68±2 |

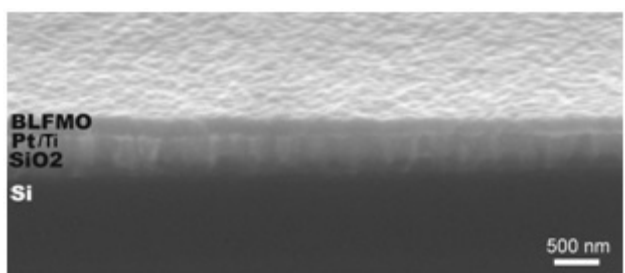

Figure 1 Typical cross section SEM image of the as-grown film at 550ºC.

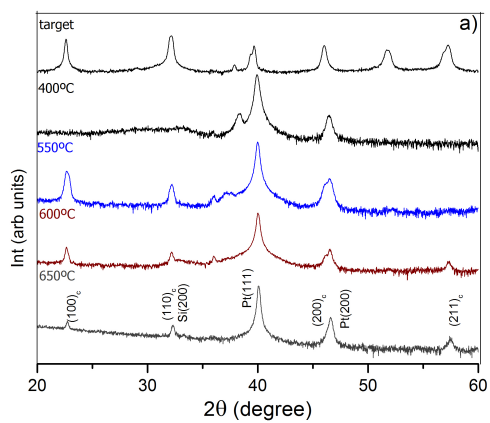

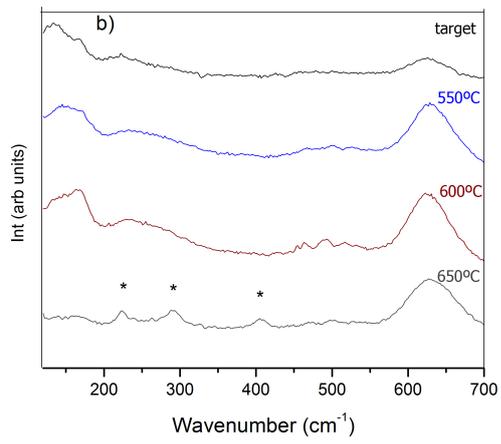

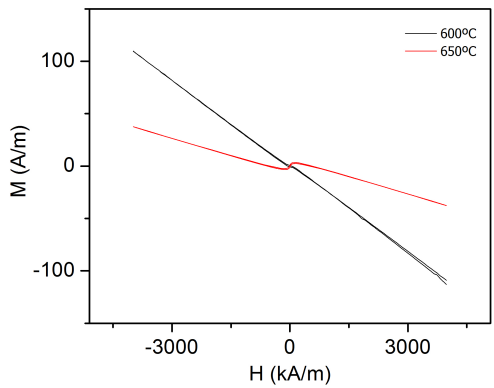

**Figure 2** a) Room temperature X-ray patterns of thin films as-grown at different temperatures. Peaks signalized with a (C) are assigned to the pseudo-cubic BLFM phase; b) Room temperature Raman spectra of thin films as-grown at different temperatures; Modes of $Fe_2O_3$ are indicated with * symbol; c) Magnetization as a function of the magnetic field strength, recorded at room temperature, for the as-prepared films at 600ºC and 650ºC.

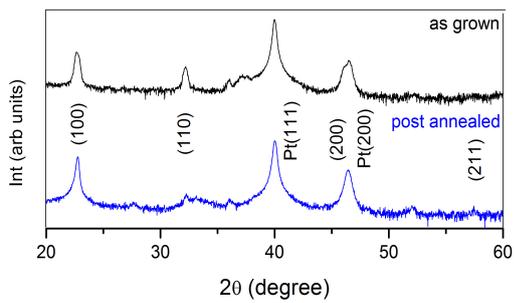

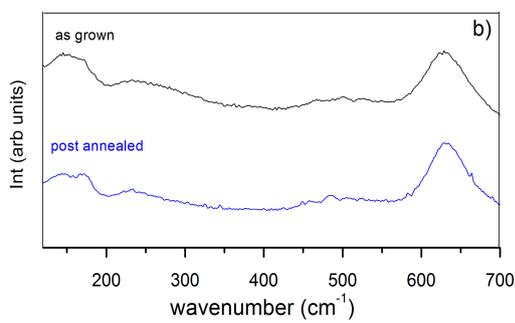

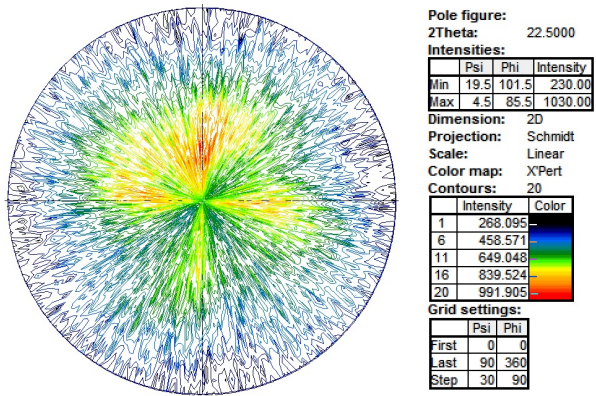

**Figure 3** Room temperature X-ray patterns (a), and room temperature Raman spectra (b) of the as-prepared film at 400ºC, post annealed at 575ºC, and the as-grown film at 600ºC; c) Pole figure of film post annealed

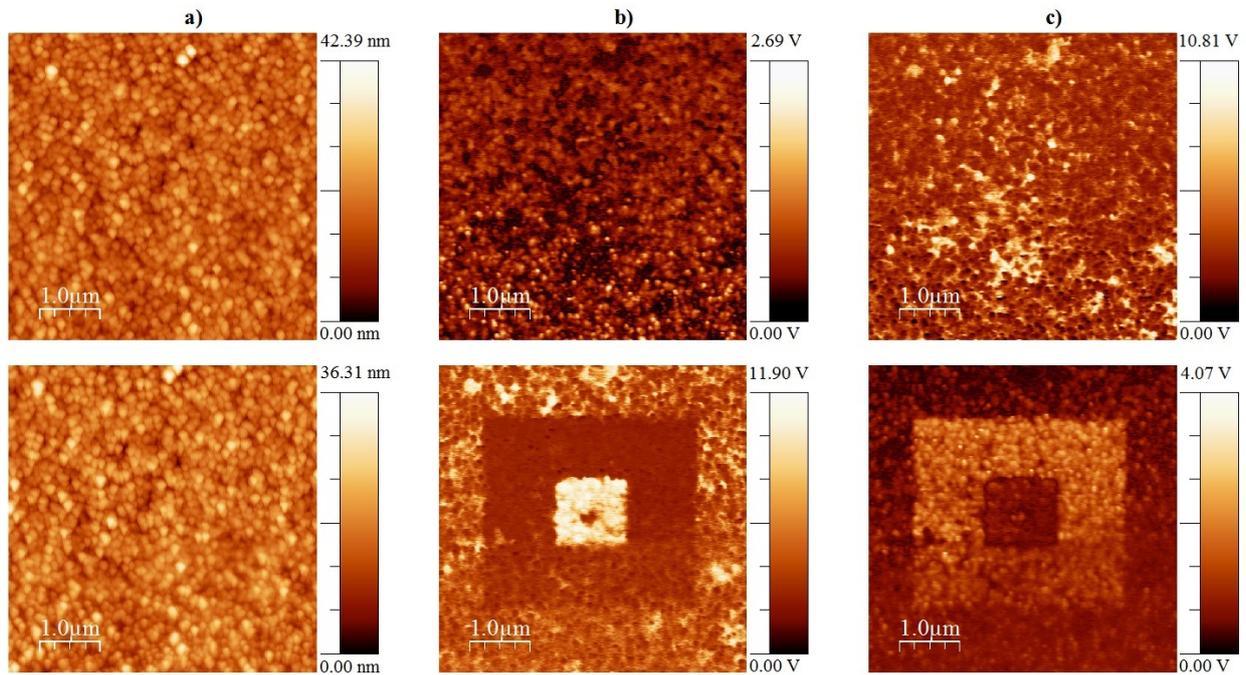

**Figure 4:** 5x5μm PFM scans of the as-prepared film at 400ºC, post annealed at 575ºC: (**a**) topography, (**b**) amplitude and (**c**) phase, before (top row) and after (bottom row) lithographic bias poling at + and − 10 V$_{dc}$.